\documentclass[11pt]{article}

\usepackage{times}
\usepackage{amssymb}
\usepackage{amsmath}
\usepackage{fancyvrb}
\usepackage{latexsym}
\usepackage{stmaryrd}
\usepackage{soul}
\usepackage{graphicx}
\usepackage{color}
\usepackage{url}
\usepackage[numbers]{natbib}
\usepackage{setspace}
\usepackage{booktabs} 
\usepackage{multirow}
\usepackage{pifont}
\usepackage{fancyvrb} 

\newtheorem{definition}{Definition}[section]
\newtheorem{axiom}[definition]{Axiom}

\newtheorem{theorem}[definition]{Theorem}
\newtheorem{remark}[definition]{Remark}
\newtheorem{corollary}[definition]{Corollary}

\newtheorem{example}[definition]{Example}

\newcommand{\NN}{\mathbb{N}}
\newcommand{\RR}{\mathbb{R}}

\newcommand{\power}{{\mathcal P}}

\newcommand{\Tquotient}{T X /\!\! \simeq\,}

\newcommand{\sem}[1]{[\![#1]\!]}

\newcommand{\lift}[1]{\{\!\!\{#1\}\!\!\}}

\newcommand{\TList}{{T_{{\sf List}}}}

\newcommand{\TTree}{{T_{{\sf Tree}}}}

\newcommand{\Node}{{\sf N}}
\newcommand{\Branch}{{\sf B}}

\newcommand{\ival}{\mathrm{iv}}
\newcommand{\spro}{\mathrm{sp}}

\newcommand{\EXT}{{\bf EXT}}
\newcommand{\INT}{{\bf INT}}

\newcommand{\SIND}{{\bf IND}}

\newcommand{\Con}{{\sf C}}

\begin{document}

\onehalfspacing

\title{On Rational Choice and the \\ Representation of Decision
  Problems\footnote{We thank Thomas Epper, three anonymous refrees, and seminar audiences at the University of St. Gallen and the European Meeting of the Econometric Society 2019 for helpful comments.}}

\author{Paulo Oliva \\
\footnotesize School of Electronic Engineering and Computer Science, Queen Mary University London
\vspace{0.01cm}\\
Philipp Zahn\footnote{Corresponding author. Email: philipp.zahn@unisg.ch. Postal
address: Varnbueelstrasse 19, CH-9000 St.Gallen, Switzerland} \\
\footnotesize Department of Economics, University of St. Gallen \\
}

\maketitle

\begin{abstract}
 In economic theory, an agent chooses from available alternatives -- modeled as a \emph{set}. In
 decisions in the field or in the lab, however, agents do not have access to the
 set of alternatives at once. Instead, alternatives are represented by the
 outside world in a structured way. Online search results are \emph{lists} of
 items, wine menus are often \emph{lists of lists} (grouped by type or country),
 and online shopping often involves filtering items which can be viewed as
 navigating a \emph{tree}. Representations constrain how an agent can choose. At
 the same time, an agent can also leverage representations when choosing,
 simplifying his/her choice process. For instance, in the case of a list he or she can use
 the order in which alternatives are represented to make his/her choice.

 In this paper, we model
 representations and decision procedures operating on them. We show that choice
 procedures are related to classical choice functions by a canonical mapping.
 Using this mapping, we can ask whether properties of choice functions can
 be lifted onto the choice procedures which induce them. We focus on the
 obvious benchmark: rational choice. We
 fully characterize choice procedures which can be rationalized by a strict
 preference relation for \emph{general} representations including
 lists, list of lists, trees and others.
 Our framework can thereby be used as the basis for new tests of rational
 behavior.

 Classical choice theory operates on very limited information,
 typically budgets or menus and final choices. This is in stark contrast to the
 vast amount of data that specifically web companies collect about their users'
 choice process. Our framework offers a way to integrate such data into economic
 choice models.


\end{abstract}



\newpage

\section{Introduction}

Suppose you want to buy a product from an online retailer. A search on their website will
return a page containing a list of 10 alternatives, plus the option to explore the next page containing 10 more
alternatives, and so on. Before you can explore item 11, you need to navigate to
the next page. Item 21 will be two clicks away from your initial search.

How does the presentation of alternatives affect your decision process? How can we
reason about such information and its effect? In the standard
economic treatment such information does not play any role. In fact, cannot play any role: There is no obvious
way how to incorporate non-choice data and information about the environment
into the classical choice framework. At the same time, such data has
informational value and economic
relevance. Many online platforms track extensive data about their customers
beyond mere choices in order to analyze their decision-making.\footnote{Researchers also increasingly study consumers'
  search behavior at a granular level (e.g
  \cite{moe2006empirical,bronnenberg2016zooming,shi2021path}) and aim to integrate
  different types of information in learning customers' preferences (e.g. \cite{farias2019learning}).}

The question is, how do we formally model such richer choice environments? In
this paper, we propose such a framework in which we model
\emph{representations} like the website above, \emph{choice procedures} operating on
them, and an \emph{extension map} which relates procedures to the classical choice approach in a canonical way.
We conceive of a
procedure as a program: Given the representation of alternatives it describes ``how'' an agent navigates through the
representation and how he or she arrives at his choice. Our framework can
describe rich choice environments and the behavior of agents interacting with
this environment. It therefore brings formal choice theory and actual observable
information of choice behavior closer together.

For a program we can also ask ``what'' it computes. It can be mapped into its input-output
behavior -- for each input we observe some output.\footnote{In theoretical computer science, the formal description of a program's
  behavior is often referred to as its \emph{denotational semantics}, compare
  e.g. \cite{winskel1993formal}.} We can do the same for a
choice procedure. The extension map relates each choice procedure to an input-output
function. As it happens, in the context of choice procedures these input-output functions are well known: they are ``choice functions''
that output a choice for each decision problem an agent faces. Hence, there is a
natural way to hide the richer description and extract the economic bare-bone
information -- if one wants to.

The formal mapping between choice procedure and actual choice also introduces
new ways of analyzing choice. We can investigate whether and how properties of
choice behavior  are related to properties of choice procedures.  In other words, in which way is ``what'' an agent
chooses related to ``how'' he or she chooses.
In economic choice theory, a central question concerns rationalizability: Can
choices observed by an agent be the outcome of the maximization of a preference
relation?\footnote{For an introduction to this question see, for instance,
  \cite{Rubinstein2019lecture}. There is a rich literature on different forms of
  rationalizability. For an overview see \cite{chambers2016revealed}.}
Thus, we begin our investigation with rational choice as the obvious benchmark:  What conditions do representations and choice procedures have to
fulfill so that choices can be rationalized? We distinguish two scenarios. First, we only impose minimal assumptions
regarding representations: we assume that the agent only knows the shape of a representation. For instance, an agent knows that alternatives are organized in a tree
but does not know which alternative is positioned where in the tree. We show, an agent with a
strict preference relation must have a procedure which (i) ignores \emph{any}
representation and (ii) proceeds by a 'divide-and-conquer' strategy. The
latter means that the overall choice will be determined by his/her choice on
'sub-problems'. Where do these sub-problems come from? They are determined by the
representation! In the initial example above, an agent proceeds by divide-and-conquer if
his/her ultimate choice is either among the choice on the first ten items or is the same as the
choice on the remaining result pages.

The result that a procedure has to follow the
principle of ``divide-and-conquer'' mirrors Plott's ``path-independence''
property \citep{Plott1973} and its role in rationalizable choice functions.
While these notions appear similar, and may be obvious in hindsight, there is still a subtle but important difference. Path-independence is a condition imposed on choice \emph{behavior}.
That is, if we observe a sequence of choices, we can (in principle) tell whether this
condition is violated or not.

In contrast, our property operates on the level of the
procedure, which is an operational description of how the agent will choose. As a
consequence, if an agent describes how he or she will proceed, we know whether this
condition is fulfilled before a single choice is made. This shift of perspective
and its underlying methodology originates in computer science where an analogous
problem exists.  Computer scientists are interested in the relationship between
program description and their behavior (``semantics'') as they would like to understand the behavior of the
programs even \emph{before} running them. Such understanding contributes to the
design of programs that behave as intended. The alternative to this is \emph{testing}, i.e. running the programs
on a range of inputs and observing how they behave.

In decision theory, when we think about rationalizable choices, in effect we are thinking about
testing. Given budgets, we observe choices. If these choices are consistent, the
agent chose rationally. Here, we are lifting the rationality criterion
onto the ``program'', the agent's decision procedure, and the representation of
alternatives. We ask, what structure must programs fulfill so that from its
description we can tell whether choices are rationalizable.
One might be tempted
to view this latter point as a technicality but we posit that having a way to express
decision procedures as well as being able to check properties of their induced
behavior  will
become the more relevant the more software systems make decisions on behalf of
humans. When entrusting software with making  consumption decisions, we
 want to make sure that the software makes good decisions \emph{before}
we let it loose.

Assuming that \emph{nothing} is known about
representations except their structure is an important edge case but arguable
not the most interesting one. Therefore, we turn to a second, more realistic
scenario  where information about the positioning of specific alternatives in a
representation is known.

We show how such additional information can be modelled in our setting and how
it affects the possibility to rationalize choice. If the
environment carries relevant information, rationalizable choice procedures
exist which are merely operating on the shape of a representation.

To make it more concrete: Consider an agent buying a mattress online. After filtering the right size, an agent is
presented with a list of alternatives.
The order in the list is not random but follows some criterion, say mattresses
ordered from hard to soft. If the agent's preferences
are aligned with this criterion, he or she can just choose
the first (or the last) element in the list -- even without inspecting this
alternative before.

But we can consider this aspect also from a different angle. Online environments are designed with intentions about the choice
process of customers. We can ask how a choice environment should be
structured such that agents can replace a complicated procedure with a
considerably simpler one that gives them the same choice.  Our framework allows to pin point under which condition a
given representation helps the agent to use a procedure which does not inspect
all alternatives.\footnote{Naturally, this is neither the only concern in the design
of such environments nor will such environments be designed for rational
consumers only. But our framework helps to understand how an environment may
help to reduce frictions on the consumer end. And companies which offer a simple and
effective choice environment may have a competitive edge. In
\cite{de2018consumer}, for instance, many customers searching for books online exhibit a
preference for a platform that cannot be explained by price differentials.}

While in this paper we focus our characterization of procedures on rational choice, our framework can be applied to other decision criteria different from maximization of a preference relation.
Consider satisficing \citep{Simon1956rational}: an agent distinguishes alternatives by either good-enough
or not good-enough. If alternatives are organized in a list, the agent will
search through the list and stop as soon as he or she finds a satisfiable element. Now
suppose alternatives are organized in a tree and properties of the alternatives
determine their position in the tree. For instance, in the case of a mattress this
could be size, spring- vs foam-core etc. If the agent's set of satisfiable
elements share this property, he or she can filter alternatives (i.e. go along the
branches of the tree) and just choose any element from the remaining
alternatives. The framework we propose can accommodate such alternatives to
rational choice. We illustrate this for satisficing.

The paper is organized as follows. In the next section we relate our paper to
the literature. The two sections after that then set the scene:
In Section \ref{sec-representation}, we
introduce the notion of a represented decision problem, and relate it to classical
decision problems. In Section \ref{subsec:procedures} we describe decision
procedures on these represented problems, and relate it to choice functions. We
then come to the first result: In Section \ref{sec:rational_choice} we introduce properties of decision procedures and use
them to characterize rationalizable procedures.
The second result follows in Section \ref{sec:meaning} where we consider additional restrictions of
representations so that rational agents can rely on them when choosing.
Section~\ref{sec:conclusion} concludes.

The challenge in this paper is setting up the framework, i.e. adequate
definitions. The main theorems then fall out in a straightforward way.
But we consider this a feature and not a bug. As the proofs proceed in a standard fashion, we have
delegated them to the Appendix.

\section{Related literature}\label{sec:literature}

The general question how the information about the environment can influence choice is, of course, not a novel idea.
The role of the external environment and its interaction with internal choice processes
was already discussed in \cite{Simon1956rational}.\footnote{For a more recent account see
  \cite{Todd2003}.} In this paper, we propose a way to model the
interaction between external information and internal choice processes and
analyze their properties.

Our paper is related to several recent strands of the literature. The paper closest to
ours is \cite{Salant2008} who also investigate the role of extra information on
choice behavior. They provide a model which extends a classical
choice problem to a tuple containing a budget and ``frame''. They then investigate how choices by agents who make use of the frames can be related to rational choice.
There are two differences to our paper: First, instead of assuming ``(...) that the frame affects choice
only as a result of procedural or psychological factors'' so that ``additional
information that is in fact relevant in the rational assessment of alternatives
thus should not be regarded as frame'' as in
\cite{Salant2008}[p.1288], we focus on the opposite: how can the extra
information support rational decision-making. Secondly, our approach differs. In
\cite{Salant2008} properties and characterizations are in terms of
behavior -- analogously to classical choice theory. In our framework, properties
and characterizations are in terms of the description of how agents choose. This
changes the interpretation of properties and how they can be used.

Apart from \cite{Salant2008} there are several other papers which consider
specialized representations and then investigate choice
functions focused on them. For instance, \cite{Rubinstein2006b} investigate procedures
which operate on \emph{lists} of alternatives.\footnote{For other papers
  investigating lists see \cite{Yildiz2016,
    dimitrov2016divide}. \cite{mukherjee2014choice} considers \emph{tree-based}
representations.} In contrast, we consider representations in general and our
results hold not only for specific representations. We thereby can generalize
their results.

In this paper, we focus on the interaction between representations and choice
procedures which result in rational choice. However, representations can be used
in general to simplify procedures which in the end implement complex goals --
rational or not. This wider perspective of considering the interaction between representations
and choice procedures, for which we provide a general model, links our paper to two further strands
of the literature.
First, several papers consider sequential choice procedures where agents reduce
the overall set of alternatives in steps towards a final choice. For instance, \cite{Mandler2012} show that a heuristic,
choosing by checklist, where agents face a decision problem and sequentially
reduce the set of alternatives by throwing out alternatives which do not satisfy
desirable properties, can lead to rational choices. Representations of
alternatives can simplify this approach:
If alternatives are organized in a binary tree structure reflecting
characteristics of the goods, then the agent can use this information to
implement his/her checklist. Other sequential procedures which could be supported by representations are analyzed in \cite{Apesteguia2013,Manzini2007, manzini2012choice}. In all these cases, equipped with a suitable representation,
agents can operate simple procedures which will mimic their internal choice processes.

Second, standard theory assumes that agents have access to all alternatives at once.
However, in practice agents often do have to search for alternatives. Starting
with \cite{stigler1961economics} there is an old literature investigating the
consequences of search costs, typically focused on finding the best price. It is
obvious how a representation of alternatives will facilitate search -- just
consider alternatives arranged in a list ordered by increasing prices.

More recently, search has been investigated from the perspective of bounded
rationality and behavioral economics. For instance, \cite{caplin2019rational}
study the consequences of rational inattention on agents. They derive agents'
optimal ``consideration sets'', i.e the set of alternatives that agents will
consider at all when making a choice. Consideration sets are also analyzed in
\cite{masatlioglu2013choice}. They provide a model of how an agent's
consideration set dynamically evolves. Again, it is obvious that the
representation of goods may be interwoven with agents' consideration sets (as
\cite{masatlioglu2013choice} also point out). This could be because of the
arrangement of goods by quality (average recommendations) or popularity (most
frequently bought). Representations also matter when it comes to
understanding how consumers navigate the vast number of varieties they nowadays face. In
\cite{lleras2017more} consumers will inspect alternatives \emph{only if} they are members of
their consideration set.  Representations may help to quickly reduce the initial very
large set of alternatives.\footnote{If representations are well designed, then a
shop can also offer a larger variety of alternatives without overwhelming consumers.} Once sufficiently reduced, agents maximize
rationally. \cite{lleras2017more} provide an example of such a procedure,
``Narrowing Down'', where the agent reduces alternatives by repeatedly refining
his search up to a point where the number of listed alternatives is below a
threshold.

\section{Representing Decision Problems}
\label{sec-representation}

Classically, a decision problem is modeled as a subset $A$ of a set of alternatives
$X$, i.e. $A \subseteq X$ and the choice function $c \colon \power{(X)} \to X$ as a function operating on the
decision problem, using $\power{(X)}$ to denote the power-set\footnote{We will
  always assume that $A \subseteq X$ is non-empty. } of $X$ (see
Chapter 2 in \cite{kreps2018notes} or Chapter 1 in \cite{Mas-Colell1995}). This characterizes the behavior of an agent.

The modeling of a decision problem as a \emph{set} abstracts from how the
problem is presented (see \cite{Rubinstein2019lecture} pp. 24 for a discussion).
The \emph{choice function} models an agent's \emph{behavior} mapping budgets into
choices. It thus abstracts from the actual process of choosing. And indeed the
behavior of different decision-making models can be captured through
choice functions. For instance, the behavior of the satisificing procedure can
be modelled in this way (see Chapter 3 in \cite{Rubinstein2019lecture}
for a discussion).

In practice, however, agents do not see the set of alternatives (at once), but
rather a representation of that set (cf. \cite{Rubinstein2006b,Salant2008} and
the related research discussed in Section \ref{sec:literature}).
Consequently, the process of choosing involves navigating this represented problem.
For instance, some online retailers only offer a single alternative at a time. To access more options, agents have to
navigate to the next item. Or some online shops will first ask
specific questions one at a time, and depending on the answers they will offer a single alternative.
In this section, we propose a model for these ``access restrictions'', by
introducing the formal notion of a representation space.

\subsection{Representation spaces}
\label{sub-sec-dec-prob}

Before we proceed with the formal definition of a \emph{problem representation}
and of a \emph{representation space}, let us outline two properties representations should have:
\begin{itemize}
	\item \emph{Representations should be inductively defined}: We want our
    problem representations to be \emph{inductively defined}, i.e. the
    representation of large decision problems should be built from the
    representation of some of its sub-problems. For instance, a large list might
    be formed by concatenating two smaller lists, or a decision tree should be built from smaller decision trees.
    This is an important feature, as it ensures that represented problems are constructed in a \emph{modular} way.
	\item \emph{Representations should be parametric on the set of alternatives
      $X$}: Similar to the definition of a decision problem which is valid for
    different kinds of alternatives we also want to define representations without being tied up to a particular set of alternatives $X$. For instance, we should be able to define the ``list representation" of $X$, without referring to anything specific about $X$, so that we can then deal with ``lists of cars" or ``lists of wines" in a uniform way. This is also an important feature, as it ensures that representations are constructed in a \emph{uniform} way.
\end{itemize}
Recall, in the introduction we used the example of shopping online. Products $X$
are displayed as a paginated list of results, with 10 results on each page. This
representation is \emph{inductively constructed} from the set $X$ via a limited number
of operations: creating a single item, organizing 10 items in a page, and
finally combining all pages into a list. Moreover, the representation structure is also \emph{uniform} in $X$,
we can use the same structure for different sets of alternative $X$. Like mathematical expressions built
from numbers via mathematical operations, or like sentences built from words
which are themselves built from letters, search results are built from primitive
objects and operations defined on them.\footnote{In mathematics, such structured
  sets are \emph{algebras}. They have led to the development of \emph{algebraic
    types} in computer science on which we base our approach. See Chapters on Finite Data Types (Product Types and Sum Types) in \cite{Harper2016}.}

In other words, we can think of our search results as a formal language that uses the values of $X$ in some fixed way to build lists of pages of items. Formally, this boils down to thinking of represented decision problems as \emph{words in a grammar}. For instance, the grammar that describes a search result in our online retailer would be:
\begin{itemize}
	\item[] Item $\Rightarrow$ $x$, for each $x \in X$
	\item[] Page $\Rightarrow$ [Item$_1$, $\ldots$, Item$_{10}$]
	\item[] List $\Rightarrow$ Page \, or \, Page, List
\end{itemize}
The first line says that any $x \in X$ is considered a ``Search Item''. The second line says that a
list of 10 items forms a ``Page''. And the last line says that a ``List of Results'' is inductively
defined as either a single page, or an initial page followed by more pages. It
is this last recursive definition (we are defining List in terms of List) that
allows us to describe lists of arbitrary (and possibly infinite) length in a
finite way.

\begin{figure}[h]
\begin{center}
\includegraphics[width=5cm]{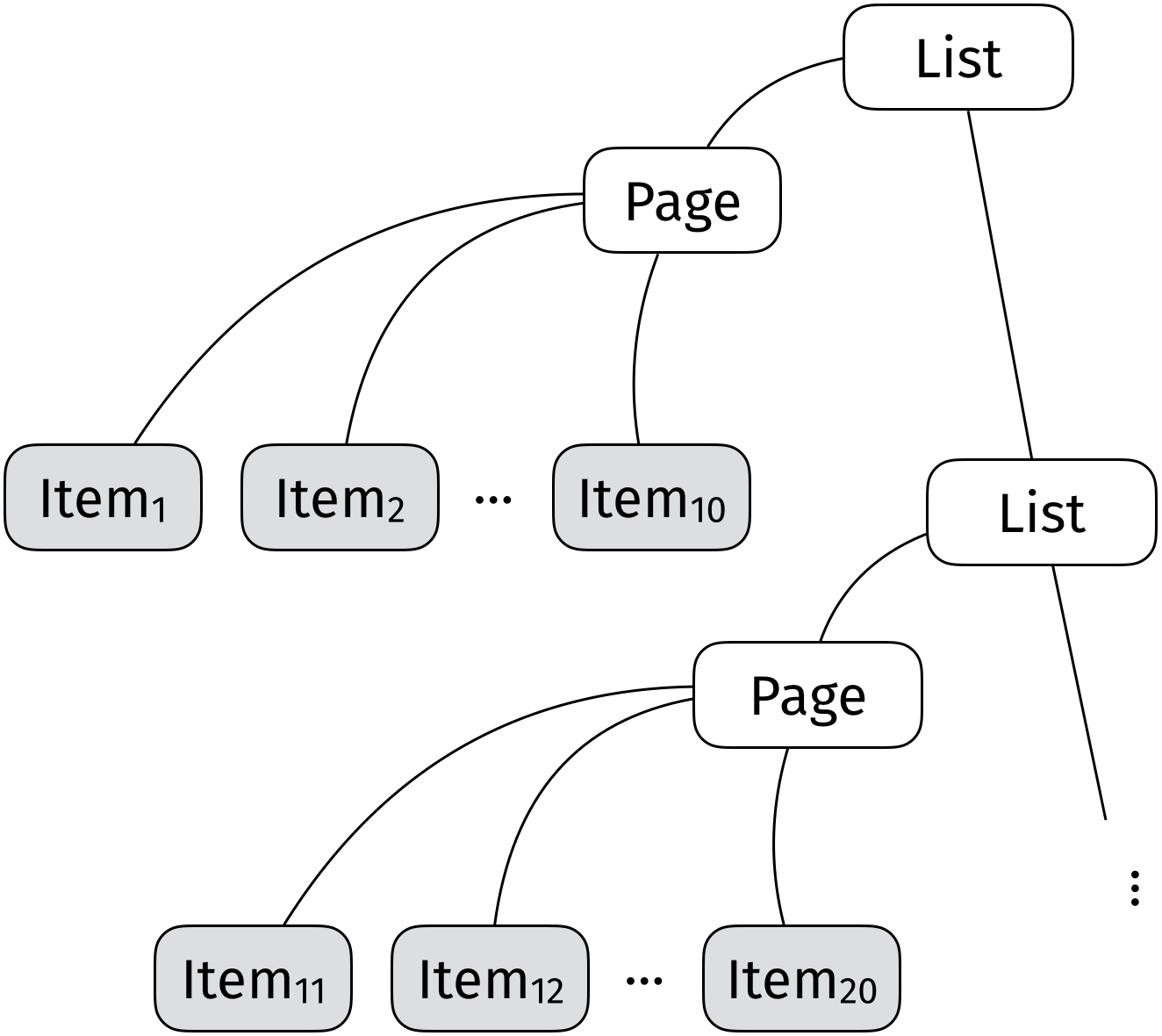}
\end{center}
\caption{Diagrammatic representation of search result}
\label{tree search}
\end{figure}

The above representation of a search result fulfills our two desiderata -- it is
both \emph{inductively defined} and \emph{parametric on the set of alternatives
  $X$}. Moreover, notice that any concrete search result can be presented diagrammatically as shown in Figure \ref{tree search}, where the various items are placed at the terminal nodes, and the internal nodes are used to describe the various labels such as ``Page'' or ``List''. A user choosing one of the items in a search result needs to navigate this structure in order to reach each of the items available.

Let us move on to a formal definition:

\begin{definition}[$T$-representation for $2^X$] \label{basic-ADT-def} Given a
  set of triples $T = \{ \langle \Con_i, n_i, m_i \rangle \}_{i \in I}$, where
  each $\Con_i$ is a name and $n_i, m_i \geq 0$ are natural numbers, with $\Con_i \neq \Con_j$ for $i \neq j$, let $T X$ denote the set obtained by repeated application of the following rule
\begin{itemize}
	\item[] If $\{ x_1, \ldots, x_{n_i} \} \subseteq X$ and $\{ a_1, \ldots, a_{m_i} \} \subseteq T X$ then $\Con_i \, x_1 \ldots x_{n_i} \, a_1 \ldots a_{m_i} \in T X$, for each $i \in I$.
\end{itemize}
We call $T X$ the \emph{$T$-representation space of $2^X$}, and each $a \in T X$ a \emph{$T$-represented decision problem}. We also refer to each name $\Con_i$ as a \emph{constructor}, as these are the only possible ways of creating represented problems in $T X$.
\end{definition}

When constructing new elements of $T X$, we can use both, elements of $X$ (the $x_1, \ldots, x_{n_i}$) or previously constructed elements of $T X$ (the $a_1, \ldots, a_{m_i}$). So $n_i$ and $m_i$ specify the arity of the constructor $\Con_i$ on each of these.

\begin{remark}
  We must have at least one $\Con_i$ with $n_i > 0$ and $m_i = 0$, to ensure that $T X$ is not an empty set. We will assume this is the case from now on.
\end{remark}

Note, the diagram above might suggest that we are restricted to representations of
decision problems as trees. But the notion of a T-representation is
more general and includes other representations of alternatives such as lists, queues,
stacks, etc.\footnote{One can think of a T-representation as \emph{sentences in a
  language}. In any language which is described by a grammar (which includes
both computer programs but also natural language) one can "parse" the sentences
of that language into a syntax tree. But that syntax tree representation of
sentences is in no way restricting the expressive power of languages themselves.
Once we fix a T-representation we are  in some sense fixing the grammar of the language, and the diagrammatic
representation of the elements in this language is nothing more than the syntax
tree of that element of the language.}

Let us illustrate the definition\footnote{In functional programming these $T$-representations are known as
  \emph{algebraic data types} (Chapter 8, \cite{Hutton2016}). For instance,
  $\TList = \{ \langle {\sf S}, 1, 0 \rangle, \langle {\sf C}, 1, 1 \rangle \}$
  corresponds to the data type of non-empty lists.}\label{fn:ADT} above using our search result example. In this case, we are actually
using three nested representations. First, single items need to be represented,
which we do with $T_{{\sf Item}} = \{ \langle {\sf I}, 1, 0 \rangle \}$. Here we are using the name ${\sf I}$ to label an item, $n = 1$ (we use one element of $X$) but $m = 0$ (we don't use previously created items). That only allows us to represent singleton sets. For instance, if $X = \{ x, y, z \}$, then $T_{{\sf Item}} X$ only has three elements
\[ T_{{\sf Item}} X = \{ {\sf I} \, x, {\sf I} \, y, {\sf I} \, z \} \]
We think of ${\sf I} \, x$ as representing the set $\{ x \}$ as an ``item" of a search result. So, the representation space $T_{{\sf Item}} X$ allows us to represent the three singleton sets $\{ x \}, \{ y \}$ and $\{ z \}$.

But we can then group items into a page -- we think of a page as a list of 10 represented items. Formally, we would take $T_{{\sf Page}} = \{ \langle {\sf P}, 10, 0 \rangle \}$. Here we use the name ${\sf P}$ to label a page, $n = 10$ (we use 10 elements to create a page) but $m = 0$ (we do not use previously created pages to create a new page). Hence, $T_{{\sf Item}} \, Y$ is the set of ``pages" each containing 10 elements from the set $Y$. If we take the set $Y = T_{{\sf Item}} \, X$, i.e. items from $X$, we are then nesting representations: a page represents a list of 10 represented items. For example, given $x_1, \ldots, x_{10} \in X$, one possible element of $T_{{\sf Page}} (T_{{\sf Item}} X)$ is
\[ {\sf P} \, ({\sf I} \, x_1) \ldots ({\sf I} \, x_{10}) \]
which we view as a representation of the set $\{ x_1, \ldots, x_{10} \}$. We are not ruling out repetitions in the represented problem. So we could have that all $x_i$ are equal to some $x$, which means that this page with apparently 10 items actually represents the singleton set $\{ x \}$. Also note that the order in which the elements appear in the representation matters. The representation where the elements are listed in inverse order
\[ {\sf P} \, ({\sf I} \, x_{10}) \ldots ({\sf I} \, x_1) \]
is a different representation of the same set $\{ x_1, \ldots, x_{10} \}$.

Finally, the representation of the ultimate search result as a list of pages involves a choice: a list is either a single page, or a page followed by other pages. This is formally achieved by defining a representation space with two constructors $\TList = \{ \langle {\sf S}, 1, 0 \rangle, \langle {\sf C}, 1, 1 \rangle \}$ for singleton lists and compound lists. Our search result over a set of alternatives $X$ is then represented as an
element of $\TList (T_{{\sf Page}} (T_{{\sf Item}} X))$, i.e. a list of pages
that each contain 10 items.

In general, a representation models how an agent can access information
regarding alternatives in
a fine-grained manner. Which representation makes sense, depends
on the situation.

Obviously, it would be a rather futile exercise to fix
a specific ``grammar'' such as the example above and characterize when a procedure operating on it
can be rationalized. Instead, we consider grammars in the abstract. The
properties we will later introduce in order to characterize a decision
procedure,  hold for grammars in general and not just specific cases.

Our framework is sufficiently expressive so that we could study procedures
which do not pick elements of $X$ but instead represented problems such as a list
of alternatives. While this is
certainly interesting, for the beginning we believe it is more relevant to understand when a procedure results in a concrete element.

Lastly, note, as is standard in the
literature, we assume throughout that the set of alternatives $X$ is
known.\footnote{As a consequence, the agents face no uncertainty regarding the
  alternatives they choose. We keep this assumption intact in order to make our
  results comparable to the standard in the revealed preference literature.
  However, in principle, the case where the agent faces uncertainty is of course
  interesting and could be modelled within our framework.}

\subsection{The extension and representation maps}

What is the relationship between represented decision problems and classical
decision problems?

As representation spaces are inductively defined from a finite set of constructors $T = \{ \langle \Con_i, n_i, m_i \rangle \}_{i \in I}$, we can also ``deconstruct'' a represented problem $a \in T X$ in an inductive way. For instance, given a represented problem $a \in T X$ we can distinguish between the elements of $X$ which are ``immediately accessible" and those which are part of some ``sub-problem":

\begin{definition}[Immediate values and sub-problems] Given a representation $T = \{ \langle \Con_i, n_i, m_i \rangle \}_{i \in I}$ and a set $X$, define two functions $\ival \colon T X \to 2^X$ (immediate values) and $\spro \colon T X \to 2^{T X}$ (sub-problems), inductively as:
\begin{itemize}
	\item $\ival(\Con_i \, x_1 \ldots x_{n_i} \, a_1 \ldots a_{m_i}) = \{ x_1, \ldots, x_{n_i} \}$
	\item $\spro(\Con_i \, x_1 \ldots x_{n_i} \, a_1 \ldots a_{m_i}) = \{ a_1, \ldots, a_{m_i} \}$
\end{itemize}
\end{definition}

The function $\ival$ extracts the immediate values contained in that represented problem (and ignores nested elements deeper
down the representation), whereas $\spro$ performs the dual functionality, it
gives us the set of sub-problems in a given representation, ignoring the
immediate values. Using these we can define the \emph{extension} of a represented problem $a \in T X$ as the subset $A \subseteq X$ whose elements are represented in $a$:

\begin{definition}[Extension map] Given a representation $T = \{ \langle \Con_i, n_i, m_i \rangle \}_{i \in I}$, define (uniformly in $X$) the \emph{extension map} $\sem{\cdot} \in T X \to 2^X$ inductively as follows:
\[ \sem{a} = \ival(a) \cup \bigcup_{b \in \spro(a)} \sem{b} \]
\end{definition}

This is well-defined on well-founded (i.e. finite) represented problems $a \in T X$. So the mapping $\sem{\cdot}$ translates represented decision problems into their set extension. For instance, if $X$ is the set of numbers then $\TList X$ is the space of representations of lists of numbers. Using this representation, the list $[1,2,2]$ corresponds to the element $a = {\sf C} \, 1 \, ({\sf C} \, 2 \, ({\sf S} \, 2))$ which has set extension $\{ 1, 2 \}$
\[
\begin{array}{lcl}
\sem{ {\sf C} \, 1 \, ({\sf C} \, 1 \, ({\sf S} \, 2)) }
	& = & \{ 1 \} \cup \sem{ {\sf C} \, 1 \, ({\sf S} \, 2) } \\[2mm]
	& = & \{ 1 \} \cup \{ 1 \} \cup \sem{ {\sf S} \, 2 } \\[2mm]
	& = & \{ 1 \} \cup \{ 1 \} \cup \{ 2 \}  \\[2mm]
	& = & \{ 1, 2 \}
\end{array}
\]
Hence, the extension map ``extracts'' the classical decision problem from a
represented decision problem.

Another, hands on way to interpret the extension map is to view it from the
perspective of an agent who wants to inspect all alternatives available in an
online shop. In a such a shop, the agent might need to navigate through all the
pages manually and inspect all the items by hand. The extension map instead would
extract out all alternatives at once.\footnote{Comparable functionality exists
  for many web shops which allow users to show all results in one page. Of
  course, even there, depending on the number of alternatives, agents will need
  to scroll down the page manually and will not perceive all elements at once.}

\begin{definition}[Equivalent represented decision problems] We say that two represented decision problems $a, b \in T X$ are \emph{extensionally equivalent}, written $a \sim b$, if they represent the same set, i.e. $\sem{a} = \sem{b}$.
\end{definition}

Although $a = {\sf C} \, 1 \, ({\sf C} \, 1 \, ({\sf S} \, 2))$ and $b = {\sf C} \, 2 \, ({\sf C} \, 2 \, ({\sf S} \, 1))$ are different represented problems, they are extensionally equivalent as they both represent the set $\{ 1, 2 \}$.

Conversely, we can also consider functions that take a classical decision problem $A \subseteq X$ and represent it as an element of the representation space $T X$. Hence, a represented decision problem can be thought of as an ``enriched'' classical decision problem.

\begin{definition}[$T$-representation map] \label{def-t-representation} Let $2^X$ denote the set of non-empty finite subsets of $X$.  A map $r \colon 2^X \to T X$ will be called a \emph{$T$-representation map} if $A = \sem{r(A)}$, for any $A \subseteq X$.
\end{definition}

A $T$-representation map provides a specific way of representing a classical
problem as an element of the representation space $T X$. We assume in this paper
that the representations $T X$ admit a $T$-representation map. This is necessary
so that any decision problem $A \subseteq X$ can be represented by an element $a
\in T X$, i.e. $\sem{a} = A$. All of the standard representations such as lists
and trees admit a representation map.

The $T$-representation map can also be practically interpreted from the perspective of a web
shop. In essence, it describes how alternatives should be represented. Should
they be listed? If so, how many items can a customer view at once? How will a
customer proceed to the next results? Or, should alternatives be organized in a
tree, for instance by a configuration assistant?

\subsection{Intensional aspects of the representation}

\begin{figure}[t]
\begin{center}
\includegraphics[width=10cm]{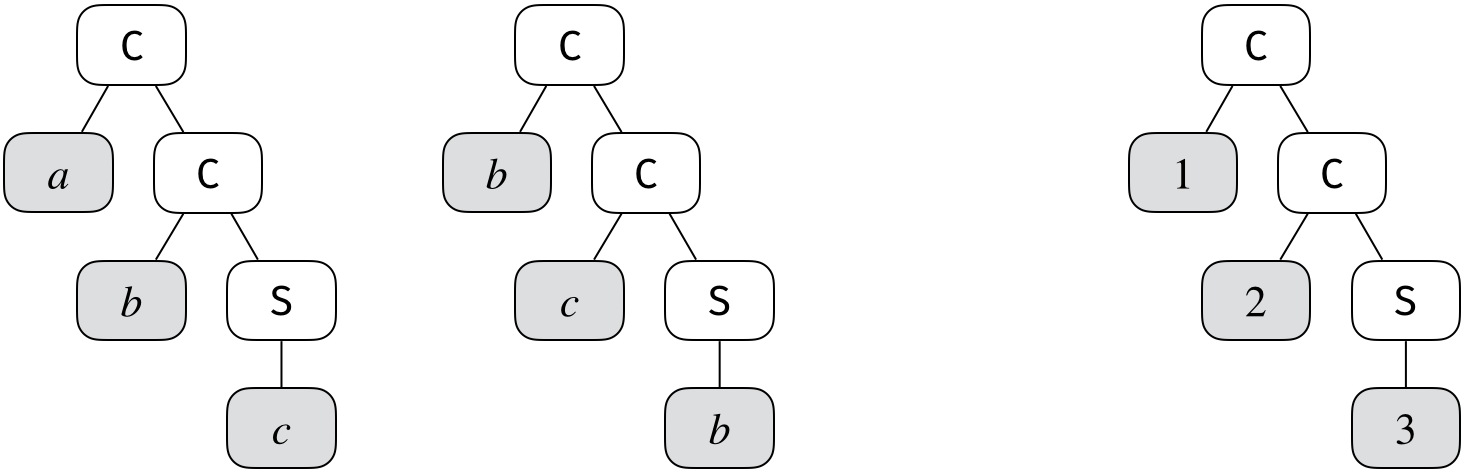}
\end{center}
\caption{Two lists over $X = \{ a, b, c \}$ (left) and their common index
  representation (right)}
\label{tree search index}
\end{figure}

Let us conclude this section discussing purely \emph{intensional} aspects of the representation, i.e. properties of the representation that are independent of the values being represented. For instance, given a represented problem $a \in T X$, we can navigate the representation in a depth-first search and inductively replace the values in $X$ by natural numbers (see Figure \ref{tree search index} where this is done for elements of $\TList \, X$ with $X = \{a, b, c \}$).

\begin{definition}[Index representation] \label{def-index} Given a representation $T = \{ \langle \Con_i, n_i, m_i \rangle \}_{i \in I}$ and a set $X$, let $\iota_X \colon T X \to T \NN$ denote the function which, given a represented problem $a \in T X$, will inductively (in a depth-first search) replace the values in $X$ by indices in $\NN$, returning an element $\iota_X(a) \in T \NN$. We call $\iota_X(a)$ the \emph{index representation} of the represented problem $a \in T X$.
\end{definition}

We can associate with each occurrence of some $x \in X$ in $a \in T X$ its corresponding index, and also the value $x \in X$ of a given index $i \in \NN$ in $a \in T X$. For example, given the list of names
\[ a = {\sf C} \, \mbox{Mary} \, ({\sf C} \, \mbox{John} \, ({\sf S} \, \mbox{Mary})) \]
its index representation is
\[ {\sf C} \, 1 \, ({\sf C} \, 2 \, ({\sf S} \, 3)) \]
so we can say the last occurrence of ``Mary" has index 3, and the value at index
2 is ``John". We will write $a[i]$ for the value at index $i$ in the decision problem $a$, so, for instance, in the example above we have $a[2] = $ John and $a[3] = $ Mary.

Looking at the index representation is a powerful way of separating the structure of the representation from the concrete set being represented.

\begin{definition}[Equivalent underlying representation] Given $a, b \in T X$, let
\[  a \simeq b \; \equiv \;  \iota(a) = \iota(b) \]
be the equivalence relation on $T X$ which identifies two represented problems that have the same underlying index representation. 
\end{definition}

For instance, in Figure \ref{tree search index} the two lists on the left have the same underlying representation, which is captured by the index representation on the right.

\begin{definition}[Representation space quotient] We will denote by $\Tquotient$ the quotient space of $T X$ consisting of the equivalence classes of $\simeq$. We use ${\bf a}, {\bf b}$ for the elements of $\Tquotient$. It is also easy to see that the mapping $\iota \colon TX \to T \NN$ can be lifted to an injection $\iota^* \colon \Tquotient \, \to T \NN$.
\end{definition}

Indeed, by the definition of $\iota \colon T X \to T \, \NN$ and the relation equivalence class $a \simeq b$, for each ${\bf a} \in \Tquotient$ all the represented problems in $a \in {\bf a}$ will map to the same index representation $a^* = \iota(a) \in T \, \NN$, so we have $\iota^*({\bf a}) = a^*$.

\section{Decision Procedures}
\label{subsec:procedures}

In the previous section we described a general approach to deal with the \emph{representation} of decision problems via the notion of a representation space $T X$ for the set of alternative $X$. Each decision problem $A \subseteq X$ may be represented in different ways as elements $a \in T X$. Similarly, we can consider different \emph{implementations} of a choice function $c \colon 2^X \to X$ as what we call a \emph{decision procedure}:

\begin{definition}[Decision procedure] A program $P \colon T X \to X$ such that
  $P(a) \in \sem{a}$, for all $a \colon T X$, is called a \emph{decision procedure}.
\end{definition}

Therefore, a decision procedure $P \colon T X \to X$ computes an element $x = P
a$ from a \emph{represented} decision problem $a \in T X$. Procedures are
defined by a case-distinction on each of the constructors of $T X$.  For each
constructor we need to state how the procedure will operate, and what the next
step in the decision process will be. In this way the representation of a decision problem influences how an agent can
choose.\footnote{Note, the references to ``program'' is not accidental. In
  computer science, a decision procedure corresponds to a functional
  program operating on an algebraic data type. In the context of a program, the
  case distinctions in the decision
  procedure are referred to as ``pattern-matching''.}

\begin{example}[First element on list] \label{example-A} Let us illustrate this point with our representation of lists over some set of alternative $X$, i.e. $\TList \, X$. Suppose the agent always chooses the first element of the list. His/her procedure $P \colon \TList \, X \to X$ can be described as
\begin{itemize}
    \item[] $P (a) = \left\{
    \begin{array}{ll}
    	x & \mbox{if $a = {\sf S} \, x$} \\[1mm]
    	x & \mbox{if $a = {\sf C} \, x \, xs$}
    \end{array}
    \right.$
\end{itemize}

Given $a = {\sf S} \, x$, a single element list, the procedure chooses this single
element $x$. Given $a = {\sf C} \, x \, xs$, a compound list containing a ``head'' element $x$ and a
``tail'' $xs$, the procedure also chooses the head. Since any element of
$\TList$ is in either one of the two forms, the procedure is a well defined function.
\end{example}

\begin{example}[Left-most element on binary tree] \label{example-D}
Consider now a representation space for \emph{binary trees} $\TTree = \{ \langle \Node, 1, 0 \rangle, \langle \Branch, 0, 2 \rangle \}$. So the elements of $\TTree \, X$ are binary tree representations of subsets of $X$. These will contain single ``node'' elements $\Node \, x$, for values $x \in X$, and internal ``branch'' elements $\Branch \, a_l \, a_r$, where $a_l, a_r$ are the ``left'' and ``right'' sub-trees. In this case, a decision procedure $P \colon \TTree \, X \to X$ on binary trees that always selects the left-most element on the tree can be defined \emph{recursively} as
\begin{itemize}
   \item[] $P(a) = \left\{
    \begin{array}{ll}
    	x & \mbox{if $a = \Node \, x$} \\[1mm]
    	P(a_l) & \mbox{if $a = \Branch \, a_l \, a_r$}
    \end{array}
    \right.$
\end{itemize}
Note how each of the two rules above describes a local decision: a decision on a final node, or a decision on an internal branch. A global choice process evolves from the iteration of these local descriptions via recursion.
\end{example}


\subsection{Relation to choice functions}

In the same way as a decision problem can have multiple representations, a choice
function can be implemented in different ways. Note, however, each choice function $c \colon 2^X \to X$ gives rise to one \emph{canonical} decision procedure: Given a represented problem $a \in T X$, look at its set extension $A = \sem{A}$ and apply the choice function to this set $x = c(\sem{a})$.

\begin{definition}[Canonical decision procedure from a choice function] \label{canonical-proc-from-choice} Given a choice function $c \colon 2^X \to X$, we can use the extension mapping $\sem{\cdot} \colon T X \to 2^X$ to lift $c \colon 2^X \to X$ into a decision procedure, which we denote by $\sem{c} \colon T X \to X$, as
\[ \sem{c}(a) = c(\sem{a}) \]
We will call $\sem{c}$ the \emph{canonical decision procedure} associated with the choice function $c \colon 2^X \to X$.
\end{definition}

But this is just one of the possible ways in which a choice function can be
``implemented'' as a decision procedure. Suppose the choice function is simply
maximizing over some strict preference relation. When implementing this choice
function as an actual procedure on the representation, one would have to decide
on how to search the represented problem, which could be done for instance as a
depth-first-search (starting from left or right) or a breadth-first-search.
Although these will all lead to the same element being chosen (assuming a strict
preference relation), the computational costs of the different search strategies may be very different for each concrete choice of representation.

Now, given a $T$-representation map $r \colon 2^X \to T X$, i.e. a fixed way of representing decision problems $A \subseteq X$, then we can also convert a decision procedure $P \colon T X \to X$ into a choice function: Given a decision problem $A \subseteq X$, first represent it as an element $r(A) \in T X$ and then apply the decision procedure $x = P(r(A))$.

\begin{definition}[Choice function from decision procedure] \label{def-choice-function} Given a decision procedure $P \colon T X \to X$ and a $T$-representation map $r \colon 2^X \to T X$, define the choice function $\lift{P}_r \colon 2^X \to X$ as
\[ \lift{P}_r(A) = P(r(A)) \]
\end{definition}

\section{Decision Procedures and Rational Choice}
\label{sec:rational_choice}

When is a \emph{decision procedure} rationalizable by a strict preference
relation? To answer this question, we begin by introducing some essential axioms of decision procedures.

\subsection{Properties of decision procedures}
\label{sec:prop_proc}

The first property captures the notion of ``divide-and-conquer'':

\begin{axiom}[Inductive procedure]    A decision procedure $P \colon T X \to X$ is said to be \emph{inductive} (\SIND{}), if the choice on a given decision problem is either one of the immediate values, or the same as a choice on one of the sub-problems: Formally, for all decision problems $a \in TX$
   \[
   \mbox{$P(a) \in \ival(a)$ or there exists $b \in \spro(a)$ such that $P(a) = P(b)$}
   \]

\end{axiom}

 An agent whose procedure fulfills \SIND{}
 decides by dividing the overall problem and choosing on
 the sub-problems. The overall choice is determined by the choices on the
 sub-problems. For instance, in a restaurant, an agent who first decides
 whether to eat fish and then decides for sea bass would follow such a strategy.

 A similar notion was introduced into decision
 theory by \cite{Plott1973} under the name path-independence. There is an
 important difference though. Plott's property is defined in terms of choice
 functions, i.e. behavior. \SIND{}, in contrast, is defined for the decision
 procedure, i.e. the description of how agents proceed.

 \SIND{}
 is also related to \emph{partition independence}, a property introduced in \cite{Rubinstein2006b} as their
 interpretation of Plott's path independence property. It is defined for choice
 functions selecting elements  from lists. As Plott's property
 it is applied in terms of behavior. If we wanted to translate their property in
 terms of choice procedures, partition independence would be a considerably
 stronger notion than \SIND{} on lists. The reason is that, in combination with their
 definition of lists, in \cite{Rubinstein2006b} agents have access to the whole
 list at once and can break it up at
 arbitrary positions. Partition independence implies that procedures are
 insensitive to where the list has been separated. The inductive property only
 implies agents apply the procedure ``consistently'' on the sub-problems. The
 latter are defined by the constructors of a representation. Thus, another way
 to view our results is that we can prove rationalizability for representations
 in general while imposing less structure than partition independence would.

The property \SIND{} focuses on how a decision procedure operates on a problem.
Next we turn to the question how a procedure makes use of representation. There
are three possibilities. Given a represented problem $a \in TX$, a procedure can
\begin{itemize}
	\item[(1)] choose to ignore the representation completely, and choose only based on the set extension of the problem $\sem{a}$ (e.g. maximizing over some strict preference relation),
	\item[(2)] choose to ignore the elements themselves completely, and choose purely based on the representation (e.g. \emph{first element on list})
	\item[(3)] or a combination of the above.
\end{itemize}

The following property captures class (1):

\begin{axiom}[Extensional procedure]
    A decision procedure is \emph{extensional} (\EXT{}) if it does not make use of the representation of the problem when making the decision, i.e. it produces the same outcome on different representations of the same decision problem:
   \[ \forall a, b \in T X (a \sim b \; \Rightarrow \; P a = P b) \]
   Recall that $a \sim b$ is defined as $\sem{a} = \sem{b}$, i.e. $a$ and $b$ are representations of the same set.
 \end{axiom}

The next definition captures class (2):

 \begin{axiom}[Intensional procedure] A decision procedure is \emph{intensional} (\INT{}) if it does not make use of the actual values of the represented decision problem, but chooses simply based on the representation structure. We can make use of the quotient space $\Tquotient$ to make this precise:
   \[ \forall {\bf a} \in \Tquotient \exists i \in \NN \forall a \in {\bf a} (P(a) = a[i]) \]
   Recall that we write $a[i]$ for the value $x \in X$ at index $i \in \NN$ in the decision problem $a \in T X$.
\end{axiom}

   One should read the above as: Given a class of represented problems that only differ by the values, but not by the underlying representation, there is one particular position or index $i$ which always contains the chosen element for all problems in that given class.

Hence, these two properties capture extreme classes of decision procedures: An
 extensional procedure strips away the representation and only considers the
 classical decision problem. That is, an agent who chooses in such a way needs
 access to all alternatives. If faced with having to navigate through all
 alternatives or getting just all alternatives at once, such an agent may prefer
 the latter.

 A prominent example of an extensional procedure is of course a
 \textbf{maximizing} agent (Example \ref{example-max}):

 \begin{example}[Maximizing] \label{example-max} Consider a procedure $P \colon T
   X \to X$ that recursively goes through a represented decision problem $a \in T
   X$ to find the maximal element according to some predetermined strict
   preference relation $(X, \succ)$, i.e. a complete, transitive and
   anti-symmetric relation. Such a procedure can be recursively defined as: For each constructor ${\sf C}_i$ of $T X$ (of arities $n_i$ and $m_i$)
   \begin{itemize}
   \item[] $P ({\sf C}_i \, x_1 \ldots x_{n_i} \, a_1 \ldots a_{m_i}) = \max_{\succ} (\{ x_1, \ldots, x_{n_i} \} \cup \{ P(a_j) \, | \, j \in \{1, \ldots, m_i\} \})$
   \end{itemize}
   Note that this procedure can operate on different representations.
 \end{example}

 An intensional procedure, in contrast, considers only
 the representation. Hence, the choice is independent of the concrete alternatives
 and dependent only on the representation. Examples of intensional procedures
 are the \textbf{first option on a list} (Example \ref{example-A}) or the \textbf{left-most option on a
   binary tree} (Example \ref{example-D}).\footnote{Examples \ref{ex:T2satis},
   \ref{example-G}, and \ref{example-H} which we describe in Appendix
   \ref{subsec:examples} belong to class (3) as they are neither fully
   extensional nor fully intensional.}

 Such examples may seem contrived at first. Why would anyone choose in such a
 way without considering the alternatives? But consider an agent who searches on
 Google and just picks the first result. Or an agent who goes on Amazon and picks the first
 recommended alternative. Or an agent who in his favorite restaurant asks the
 waiter what wines he or she would recommend for a given meal and just goes with the
 first recommendation. We posit that such simple choice procedures are part of
 how people sometimes choose. Moreover, we posit that such procedures
 particularly make sense when the representations of alternatives and the choice
 environment more generally carries information that is meaningful for agents.
 We will revisit this aspect in Section \ref{sec:meaning}.

\subsection{Rationalizable Choice Procedures}
\label{sec:prop_proc}

We begin with the standard notion of a \emph{choice function} which can be rationalized
by a strict preference relation, i.e. such choices can be described as the
outcome of maximizing a strict preference relation by an agent (see
\cite{Rubinstein2019lecture} Chapter 3).

\begin{definition}[Strictly rationalizable choice function] A choice function $c
  \colon 2^X \to X$ is \emph{strictly rationalizable} if there exists a \emph{strict preference relation} $(X, \succ)$ such that for all decision problems $A \subseteq X$
\[ c(A) = \max_{\succ} A \]
The strictness of the preference relation ensures that this maximum element is always unique.
\end{definition}

Given the canonical way of converting a choice function into a procedure
(Definition \ref{canonical-proc-from-choice}), a procedure is strictly rationalizable if it has the same input-output behavior as the canonical procedure of a rationalizable choice function. Formally:

\begin{definition}[Strictly rationalizable procedure]\label{definition:rational_proc}A
  decision procedure $P \colon T X \to X$ is said to be \emph{strictly rationalizable} if
  $P = \sem{c}$, for some strictly rationalizable choice function $c \colon 2^X \to X$, i.e.
\[ \forall a \in T X, \; P(a) = \max_{\succ} (\sem{a}) \]
for some strict preference relation $(X, \succ)$.
\end{definition}

In other words, $P$ is strictly rationalizable if it ``implements'' a choice
function whose choices agree with the maximization of a strict preference
relation.

From now on, in order to simplify terminology, we will simply refer to
``rationalizable choice'' or a
``rationalizable procedure''. Throughout the paper we have in mind the
rationalizability by a strict preference relation. Obviously, we could consider
weaker forms of rationalizability, say allowing for indifference. However, the
case of strict preferences is the most demanding and puts the most restrains on
how an agent can use representations.

The following result shows that being extensional and inductive fully characterizes rationalizable procedures:

\begin{theorem} \label{theorem:main}
A procedure $P$ is rationalizable iff $P$ is \EXT{} and \SIND{}.
\end{theorem}
Thus, in order to choose rationally, a procedure needs to (i) ignore the representation completely and (ii) needs to proceed by a divide-and-conquer strategy. Not surprisingly, Example \ref{example-max} ({\bf maximization}) fulfills both properties.

The {\bf satisficing} procedure (Example \ref{ex:satis1}) illustrates a case
where \SIND{}~holds, but \EXT~does not.

\begin{example}[$\TList$ Satisficing] \label{ex:satis1} Consider the
  \emph{satisficing} $\TList$-decision procedure whereby an agent has an
  evaluation function $u \colon X \to \RR$ and a satisficing\footnote{This
    example is based on \cite{Rubinstein2019lecture}, p.30.} threshold $u^* \in \RR$. The agent chooses the first element of the list that is above the given threshold. If there is none, he or she chooses the last element. This procedure can be recursively defined as:
  \[
    P(a) = \left\{
      \begin{array}{ll}
        x & \mbox{if $a = {\sf S} \, x$} \\[1mm]
        x & \mbox{if $a = {\sf C} \, x \, a'$ and $u(x) \geq u^*$} \\[1mm]
        P(a') & \mbox{if $a = {\sf C} \, x \, a'$ and $u(x) < u^*$}
      \end{array}
    \right.
  \]
\end{example}

It is easy to see that {\bf satisficing} is an inductive procedure. However, it does not fulfill \EXT: consider two lists $a_1, a_2$ which contain the same elements $\{x, y\}$, i.e. $\sem{a_1}=\sem{a_2} = \{x, y\}$, but suppose $x$ comes first in $a_1$, while  $y$ comes first in $a_2$. Suppose also that $u(x),u(y)>u^*$. Then, $P(a_1) = x$ and $P(a_2) = y$. Thus $P$ is not extensional and hence is not rationalizable.

\subsection{Discussion}

Overall, Theorem \ref{theorem:main} has strong implications. If rational agents
must ignore the representation of the problem completely, does that mean that
representations are meaningless? So far, we only required that representations
give the decision problem some structure. Apart from that we did not impose other restrictions.

Yet, representations are rarely arbitrary. There is typically some logic behind. The pages on a search engine are listed according to some ``page rank'', the items on a restaurant menu are organized according to categories (starter, main, dessert), the products in an online shop are displayed either by relevance, average customer rating or price, etc.

In the next section, we explore the idea that, when restricted to a
``meaningful'' set of representations, it is indeed possible for agents to choose in a purely intensional way (e.g. always choosing the first hit on a search engine) while still being rational (e.g. maximizing page rank).

\section{Generalization: Meaningful Representations}
\label{sec:meaning}

In this section, we consider representations which carry additional structure.
That is, instead of considering the complete representation space $T X$, we
might wish to restrict ourselves to some subset of ``valid'' or ``meaningful''
representations ${\cal V} \subseteq T X$. For instance, we may be interested in
lists where the order in which the elements are listed is not arbitrary but follows some rule.
Nevertheless, we want that any decision problem is still representable as an element of ${\cal V}$:

\begin{definition}[Expressive ${\cal V}$] We call ${\cal V}$ \emph{expressive} if for any subset $A \subseteq X$ there exists an $a \in {\cal V}$ such that $\sem{a} = A$. We say that ${\cal V}$ is \emph{uniquely expressive} if such $a \in {\cal V}$ is uniquely determined by $A$.
\end{definition}

All properties of decision procedures introduced above relativize to subsets of $T X$. For instance, we can define the notion of rationalizability (Definition \ref{definition:rational_proc}) relative to a set of valid representations ${\cal V} \subseteq T X$:

\begin{definition}[${\cal V}$-rationalizable procedure] Let ${\cal V} \subseteq T X$. A decision procedure $P \colon T X \to X$ is said to be ${\cal V}$-\emph{rationalizable} if there exists a rationalizable choice function $c \colon 2^X \to X$ such that
\[ \forall a \in {\cal V}, \; P(a) = c(\sem{a}) \]
\end{definition}

We can also relativize the notions of \EXT{}~and \SIND{}~to represented problems
in ${\cal V}$, and we call these ${\cal V}$-\EXT{}~and ${\cal V}$-\SIND{}.
Furthermore, one direction of Theorem \ref{theorem:main} relativizes to any expressive set ${\cal V} \subseteq T X$, i.e.

\begin{theorem}\label{theorem:vmain}
Assume ${\cal V}$ is expressive. If a procedure $P$ is ${\cal V}$-rationalizable, then it is ${\cal V}$-\EXT{} and ${\cal V}$-\SIND{}.
\end{theorem}
For the converse direction, however, we need to impose a further requirement on the set of valid representations ${\cal V}$:

\begin{definition}[$T$-closed ${\cal V}$] Let $T = \{ \langle \Con_i, n_i, m_i \rangle \}_{i \in I}$. We say that ${\cal V}$ is $T$-\emph{closed} if whenever $\{ a_1, \ldots, a_{m_i} \} \subseteq {\cal V}$ then $\Con_i \, x_1 \ldots x_{n_i} \, a_1 \ldots a_{m_i} \in {\cal V}$, for any $\{ x_1, \ldots, x_{n_i} \} \subseteq X$.
\end{definition}

\begin{theorem} \label{theorem:vmain2}
Assume ${\cal V}$ is expressive and $T$-closed. If a procedure $P$ is ${\cal V}$-\EXT{} and ${\cal V}$-\SIND{} then it is also ${\cal V}$-rationalizable.
\end{theorem}
We concluded the previous section by arguing that the {\bf satisficing} procedure of Example \ref{ex:satis1} is not rationalizable because it is not extensional.
Consider, however, the following variant of satisficing where the set of representations is restricted:

\begin{example}[Satisficing with restricted representations] \label{ex:satis2} Let $(X, <)$ be a totally ordered set. Let ${\cal V} \subseteq \TList \, X$ consist only of lists where the elements are arranged according to the order $<$, so that if $a \in {\cal V}$ and $i < j$ then $a[i] \leq a[j]$. We can consider now the same procedure as in Example \ref{ex:satis1}, but now on this restricted set of lists.
\end{example}

This restricted set of lists ${\cal V} \subseteq \TList \, X$ is both expressive
and $T$-closed. The restriction on representations also ensures that this
procedure, although not globally extensional, is extensional on the restricted
domain of lists that satisfy the global ordering $(X, <)$. Moreover, as we argued before, the satisficing procedure is inductive. Therefore, by Theorem \ref{theorem:vmain2}, the restricted satisficing procedure is ${\cal V}$-rationalizable. This fact, that the above version of satisficing can be rationalized if constrains are put on the list representation, is discussed in \cite{Rubinstein2019lecture}[p.30].

It is easy to see that for any decision problem $A \subseteq X$ there is only
one way of  representing $A$ as a list $a \in \TList \, X$ if one at the same
time insists that the elements are listed according to some total ordering $(X, <)$. In general, if decision problems have a unique representation in ${\cal V}$, then any procedure is trivially ${\cal V}$-\EXT{} (there are no two problems with different representations). The following theorem shows that the converse must also hold:

\begin{theorem} \label{thm-ue} For any expressive ${\cal V}$, it is uniquely expressive iff all procedures $P$ are ${\cal V}$-\EXT{}.
\end{theorem}

As a consequence, if all procedures -- when restricted to ${\cal V}$ -- are either rational or extensional, then ${\cal V}$ must be uniquely expressive:

\begin{corollary} \label{cor-unique} If every procedure $P$ on $T X$ is either ${\cal V}$-rational or ${\cal V}$-\EXT{}, then ${\cal V}$ must be uniquely expressive.
\end{corollary}
%

Finally, we conclude this paper by considering procedures which are \emph{purely
intensional} and ${\cal V}$-\emph{rational}. We prove that in such cases the set ${\cal V}$ must fix an
index $i$ for all problems in ${\cal V}$ with the same structure and the same
value:

\begin{theorem} \label{main-theorem2} For any expressive ${\cal V}$, if $P \colon T X \to X$ is \INT{}~and ${\cal V}$-rational then
\[ \forall {\bf a} \in \Tquotient \exists i^{\NN} \forall A \subseteq X \exists v^X \forall a \in {\cal V} \cap {\bf a} (\sem{a} = A \to a[i] = v) \]
\end{theorem}
Indeed, if an agent is choosing rationally, his/her choice has to be the same on all
$a \in {\cal V}$ which are representations of the same set $A \subseteq X$. But
if his/her procedure is also purely intensional, the agent is choosing by always
selecting a particular index on the represented problem, irrespective of the
values being represented. Such a combination is only possible when the set
${\cal V}$ ensures that for all $a \in {\cal V}$, which represent the same set
$A \subseteq X$, a particular index $i \in \NN$ always contains the same value
$v \in X$ -- which is the maximal value on some strict preference relation on $X$.
Nothing needs to be said about all other indices $j \neq i$ though.

Let us entertain the following thought experiment: Suppose a search engine comes up with a ranking of pages which perfectly matches the preferences of the users over web pages. The search engine then lists pages, for each search term, in decreasing order of page rank. The users, once aware of this, will always pick the first page on the search result (a purely intensional procedure). This is only rational because the ``best'' page is always placed on the first position. But that, paradoxically, also implies that it does not really matter what pages come at 2nd, 3rd, ... positions, as these are never chosen by these {\bf first element on list} procedures.

\section{Conclusion}
\label{sec:conclusion}


If customers shop online, they do not interact with the set of alternatives
directly but with a representation of these alternatives. In this paper, we
introduce a novel framework in which we can model these representation as well
as the decision-making procedures of agents. We show how procedures induce
\emph{choice functions} which are well established in the economic literature.


In the context of choice functions, \emph{rationalizability} is a key aspect. We
thus take this as a starting point for our analysis and characterize under which conditions procedures
operating on represented problems can be rationalized. Importantly, with
adequate guarantees on representations the agent can choose in a purely
intensional way, i.e. he or she can choose on the basis of representations alone.


Two immediate implications follow from our results. First, through the
mapping from internal procedures to choice functions, we can ``lift''
rationality criteria from choice functions to procedures. What this means: If an
agent describes how he or she is actually choosing, we can
analyze that description in our language and directly tell -- without a single
choice being made -- whether the agent is maximizing a rational preference
relation or not. Moreover, in principle, we can apply a similar strategy to
click stream data (as for instance collected by online shops). Our results could serve as the
basis for new tests of rationality assumptions not by comparing different
choices -- as is usual in the empirical literature on revealed preferences --
but instead by tracking the choice process. The same methodology may be applicable to other theories of decision-making.

Secondly, the ability to integrate observable and measurable data structures
into a model can not only be used to extend tests for existing theories but it
can also be used to develop new models of decision-making. Today, there is a big
gap between classical choice theory which operates on rudimentary data such as sets
of alternatives and choices and the actual observable data. The question is how can
one integrate such data with classical choice theory and develop new models? The representations of
alternatives we provide is a way to do so.

We noted already that the formal treatment we use here is standard in theoretical
computer science and functional programming. It is a formal language
to describe representations (algebraic data types) and procedures (programs)
operating on these representations. While being
a formal language, which one can use for mathematical reasoning as we do here in
the paper, this language can also be implemented in
modern functional programming languages.

Consider again the representation of a list:
$\TList = \{ \langle {\sf S}, 1, 0 \rangle, \langle {\sf C}, 1, 1 \rangle \}$
 and the example list
$ a = {\sf C} \, 1 \, ({\sf C} \, 2 \, ({\sf S} \, 2))$

We can implement this directly in a functional programming language such as Haskell\footnote{\url{https://www.haskell.org/}} as follows:

\begin{verbatim}
(1) data TList x = S x | C x (TList x)

(2) a = C 1 (C 2 (S 2))
\end{verbatim}

(1) gives a definition of $\TList$ (known in Haskell as a data type). Note, that
as in our formal definition the Haskell data type is parametric with respect to the exact object it
shall represent (and is therefore parameterized by the variable ``$x$''). (2) is
the representation of the list $[1,2,2]$ as an instance of that data type. The
similarity of the Haskell representation to the formal representation is not a
coincidence. The formal structure of the representations we have introduced
can be directly implemented in modern programming languages which support this
functionality (as in the example above for instance in
Haskell). As a consequence of this, all
representations we introduce here can be directly implemented, and be used
for simulation of procedures, for instance. Or one could define a representation
framework (like an online shop), translate it into the formal framework to
reason about it, or match it to data.


Regarding future work, the framework we have introduced provides an interface to
a whole new toolkit that can be applied to choice scenarios. We will sketch
two of these.

First, in this paper, we follow the notion of behavior as it is standard
textbooks: a choice of an alternative. But online shops do gather a lot more information
about customers than just a final choice. In fact, in many situations customers
may not make a buy decision in the end. Still, their navigation through a
web shop may reveal relevant information about their decision-making: Which elements do they inspect? Where
do they spend their time? Nowadays, such information is routinely collected and
analyzed. The methodology we have introduced here can handle different forms
of such ``behavior''. Recall the type of procedure we consider:

\[ P \colon T X \to X \]

We can augment the outcome type to:

\[ P \colon T X \to M X \]

where $M$ forms some algebraic structure on top of the set of alternatives. Concretely,
this could be a non buy-decision; it could be a probability distribution, i.e.
we behavior is not deterministic but probabilistic; it could also mean that we
are not tracking the final choice but the intermediate steps -- as often
measured in clickstream data.  In the theoretical computer science literature
the structure of $M$ is well studied. And the cases like probability, or
``exceptions'' or ``trace'' can be directly interpreted as the aforementioned choice
theoretic terms.

A second extension concerns the complexity of procedures, which we do not touch in
the current paper.
Yet, our paper does set up a key prerequisite for such an analysis: the distinction
between procedure and behavior. Why? Because it allows us to define the
equivalence of procedures (as resulting in the exact same behavior). Equipped
with the notion of equivalence, we can cleanly compare procedures.

This is central if we want to understand the value of different representations.
Consider the following example: You want to buy an external harddrive. You go to
an online platform and search for the available alternatives. Suppose you have
clear preferences about attributes of the harddrive, for instance what type of
disk, capacity, noise level etc. Suppose further that the platform offers decision
support: you can filter and sort alternatives according to some attributes. \emph{How}
you can find the best alternative depends on that decision support. Designed in
an adequate way, it can lead you quickly to your best choice. Designed in an
inadequate way (for instance by not providing the functionality to filter out
alternatives), you might need to inspect alternatives step by step.

Indeed, in another paper \cite{oliva2018structure}, we show under which conditions such filtering/sorting
actions are adequately designed and will be ``quicker'' than maximization. Of
course, this issue is by no means limited to
maximization. The representation of alternatives and the information available
in general affects how people can choose. Here, we introduce a framework which
can be extended in these directions.

\bibliographystyle{unsrt}
\bibliography{ext-int-slim}

\begin{thebibliography}{10}

\bibitem{moe2006empirical}
Wendy~W Moe.
\newblock An empirical two-stage choice model with varying decision rules
  applied to internet clickstream data.
\newblock {\em Journal of Marketing Research}, 43(4):680--692, 2006.

\bibitem{bronnenberg2016zooming}
Bart~J Bronnenberg, Jun~B Kim, and Carl~F Mela.
\newblock Zooming in on choice: How do consumers search for cameras online?
\newblock {\em Marketing science}, 35(5):693--712, 2016.

\bibitem{shi2021path}
Savannah~Wei Shi and Michael Trusov.
\newblock The path to click: Are you on it?
\newblock {\em Marketing Science}, 40(2):344--365, 2021.

\bibitem{farias2019learning}
Vivek~F Farias and Andrew~A Li.
\newblock Learning preferences with side information.
\newblock {\em Management Science}, 65(7):3131--3149, 2019.

\bibitem{winskel1993formal}
Glynn Winskel.
\newblock {\em The formal semantics of programming languages: an introduction}.
\newblock MIT press, 1993.

\bibitem{Rubinstein2019lecture}
Ariel Rubinstein.
\newblock {\em Lecture notes in microeconomic theory}.
\newblock Princeton University Press, 2019.

\bibitem{chambers2016revealed}
Christopher~P Chambers and Federico Echenique.
\newblock {\em Revealed preference theory}, volume~56.
\newblock Cambridge University Press, 2016.

\bibitem{Plott1973}
Charles~R. Plott.
\newblock Path independence, rationality, and social choice.
\newblock {\em Econometrica}, 41(6):1075--1091, 1973.

\bibitem{de2018consumer}
Babur De~los Santos.
\newblock Consumer search on the internet.
\newblock {\em International Journal of Industrial Organization}, 58:66--105,
  2018.

\bibitem{Simon1956rational}
Herbert~A Simon.
\newblock Rational choice and the structure of the environment.
\newblock {\em Psychological review}, 63(2):129, 1956.

\bibitem{Todd2003}
Peter~M Todd and Gerd Gigerenzer.
\newblock Bounding rationality to the world.
\newblock {\em Journal of Economic Psychology}, 24(2):143 -- 165, 2003.

\bibitem{Salant2008}
Yuval Salant and Ariel Rubinstein.
\newblock {(A,f) : Choice with frames}.
\newblock {\em Review of Economic Studies}, 75:1287--1296, 2008.

\bibitem{Rubinstein2006b}
Ariel Rubinstein and Yuval Salant.
\newblock {A model of choice from lists}.
\newblock {\em Theoretical Economics}, 1:3--17, 2006.

\bibitem{Yildiz2016}
Kemal Yildiz.
\newblock List-rationalizable choice.
\newblock {\em Theoretical Economics}, 11(2):587--599, 2016.

\bibitem{dimitrov2016divide}
Dinko Dimitrov, Saptarshi Mukherjee, and Nozomu Muto.
\newblock 'divide-and-choose' in list-based decision problems.
\newblock {\em Theory and Decision}, 81(1):17--31, 2016.

\bibitem{mukherjee2014choice}
Saptarshi Mukherjee.
\newblock Choice in ordered-tree-based decision problems.
\newblock {\em Social Choice and Welfare}, 43(2):471--496, 2014.

\bibitem{Mandler2012}
Michael Mandler, Paola Manzini, and Marco Mariotti.
\newblock {A million answers to twenty questions : Choosing by checklist}.
\newblock {\em Journal of Economic Theory}, 147(1):71--92, 2012.

\bibitem{Apesteguia2013}
Jose Apesteguia and Miguel~A Ballester.
\newblock {Choice by sequential procedures}.
\newblock {\em Games and Economic Behavior}, 77:90--99, 2013.

\bibitem{Manzini2007}
Paola Manzini and Marco Mariotti.
\newblock {Sequentially rationalizable choice}.
\newblock {\em American Economic Review}, 97(5):1824--1839, 2007.

\bibitem{manzini2012choice}
Paola Manzini and Marco Mariotti.
\newblock Choice by lexicographic semiorders.
\newblock {\em Theoretical Economics}, 7(1):1--23, 2012.

\bibitem{stigler1961economics}
George~J Stigler.
\newblock The economics of information.
\newblock {\em Journal of Political Economy}, 69(3):213--225, 1961.

\bibitem{caplin2019rational}
Andrew Caplin, Mark Dean, and John Leahy.
\newblock Rational inattention, optimal consideration sets, and stochastic
  choice.
\newblock {\em The Review of Economic Studies}, 86(3):1061--1094, 2019.

\bibitem{masatlioglu2013choice}
Yusufcan Masatlioglu and Daisuke Nakajima.
\newblock Choice by iterative search.
\newblock {\em Theoretical Economics}, 8(3):701--728, 2013.

\bibitem{lleras2017more}
Juan~Sebastian Lleras, Yusufcan Masatlioglu, Daisuke Nakajima, and Erkut~Y
  Ozbay.
\newblock When more is less: Limited consideration.
\newblock {\em Journal of Economic Theory}, 170:70--85, 2017.

\bibitem{kreps2018notes}
David Kreps.
\newblock {\em Notes on the Theory of Choice}.
\newblock Routledge, 2018.

\bibitem{Mas-Colell1995}
Andreu Mas-Colell, Michael~D. Whinston, and Jerry Green.
\newblock {\em {Microeconomic Theory}}.
\newblock Oxford University Press, 1995.

\bibitem{Harper2016}
Robert Harper.
\newblock {\em {Practical Foundations for Programming Languages}}.
\newblock Cambridge University Press, 2016.

\bibitem{Hutton2016}
Graham Hutton.
\newblock {\em Programming in Haskell}.
\newblock Cambridge University Press, 2016.

\bibitem{oliva2018structure}
Paulo Oliva and Philipp Zahn.
\newblock The structure of the environment and the complexity of rational
  choice procedures.
\newblock {\em mimeo. arXiv 1809.06766}, 2018.

\end{thebibliography}

\appendix

\section{Additional Examples}\label{subsec:examples}

In this section, we provide additional examples which are neither fully
extensional nor fully intensional.

\begin{example}[$\TTree$ Satisficing] \label{ex:T2satis} Consider also a \emph{satisficing} procedure on a binary tree, $\TTree \, X$, whereby an agent has an evaluation function $u \colon X \to \RR$ and a satisficing threshold $u^* \in \RR$, and the agent chooses the left-most element of the binary tree that is above the given threshold. If there is none, he or she chooses the right-most element. This procedure can be recursively defined as:
\[
P(a) = \left\{
	\begin{array}{ll}
		x & \mbox{if $a = \Node \, x$} \\[1mm]
		P(a_l) & \mbox{if $a = \Branch \, a_l \, a_r$ and $u(P(a_l)) \geq u^*$} \\[1mm]
		P(a_r) & \mbox{if $a = \Branch \, a_l \, a_r$ and $u(P(a_l)) < u^*$}
	\end{array}
	\right.
\]
\end{example}

\begin{example}[Conditional $\TTree$ satisficing] \label{example-G} The set of alternatives $X$ is partitioned into two sets, $K$ (known, or default) and $N$ (new). Given a problem $a \in \TTree \, X$, if more than half of the elements in $\sem{a}$ are known elements, choose the first (on a depth-first search) element of the binary tree which is in $K$. Otherwise choose the maximal element of the binary tree according to some predetermined order $(X, \succ)$ as in Example \ref{example-max}.
\end{example}

\begin{example}[Choose default on large problems] \label{example-C} Assume the agent has a default choice $x \in X$ in mind and a size threshold $N$. Consider the procedure that given $a \in T X$ returns $x$ when $x \in \sem{a}$ and $|\sem{a}| > N$, or else returns the maximal element in $\sem{a}$ according to some predetermined order $(X, \succ)$ as in Example \ref{example-max}.
\end{example}

\begin{example}[Avoiding undesirable element] \label{example-H} Consider a procedure for selecting an element from a binary tree of alternatives $a \in \TTree \, X$. The procedure will recursively operate on the left sub-tree and take that choice, except when that choice is an undesirable element $x^* \in X$ and the number of choices is large (bigger than some given threshold $N \geq 1$), in which case it disregards the left choice and works recursively on the right sub-tree. The element $x^*$ and the threshold $N$ are fixed exogenously. The procedure can be defined as:
\[
P(a) = \left\{
	\begin{array}{ll}
		x & \mbox{if $a = \Node \, x$} \\[1mm]
		P(a_l) & \mbox{if $a = \Branch \, a_l \, a_r$ and $P(a_l) \neq x^*$ or $|\sem{a_l} \cup \sem{a_r}| \leq N$} \\[1mm]
		P(a_r) & \mbox{if $a = \Branch \, a_l \, a_r$ and $P(a_l) = x^*$ and $|\sem{a_l} \cup \sem{a_r}| > N$}
	\end{array}
	\right.
\]
\end{example}

\begin{example}[Larger sub-problem bias] \label{example-bigproblem} Consider again an agent choosing on binary trees $a \in \TTree \, X$, but this time suppose they also have a size threshold $N \geq 1$. The agent always recursively chooses to work on the biggest of the two sub-trees, and maximizes when reaching a sub-problem of size smaller or equal to $N$:
\begin{itemize}
  \item[] $P (a) = \left\{
  \begin{array}{lcl}
	\max_\succ \sem{a} & & \textrm{if} \; | \sem{a} | \leq N \quad \mbox{(as in Example \ref{example-max})} \\[2mm]
  	P(a_1) & & \textrm{if} \; a = \Branch \, a_1 \, a_2 \wedge | \sem{a_1} | > | \sem{a_2} | \\[2mm]
	P(a_2) & & \textrm{if} \; a = \Branch \, a_1 \, a_2 \wedge | \sem{a_1} | \leq | \sem{a_2} |
  \end{array}
  \right.$
\end{itemize}
\end{example}

\section{Proofs}

Proof of Theorem \ref{theorem:main}:

{\bf Proof}. First, assume $P$ is rationalizable. Let $c \colon 2^X \to X$ be a rationalizable choice function such that
\begin{itemize}
	\item[(i)] $P(a) = c(\sem{a})$, for all $a \in T X$.
\end{itemize}
Let $(X, \succ)$ be the \emph{strict preference relation} underpinning $c$, i.e.
\begin{itemize}
	\item[(ii)] $c(A) = \max_{\succ} A$
\end{itemize}
In order to show \EXT{}, let $a, b \in T X$ be two represented problems such that $\sem{a} = \sem{b}$. Clearly, $c(\sem{a}) = c(\sem{b})$, and hence, by (i), $P(a) = P(b)$.  In order to show that $P$ is also \SIND{}, let $a \in TX$ be a fixed represented problem, and assume $P(a) \not\in \ival(a)$. Since $P(a) \in \sem{a}$ there must be a $b \in \spro(a)$ such that $P(a) \in \sem{b}$. By (ii), $P(a)$ is a maximal element in $\sem{a}$, and hence also in $\sem{b}$, since $\sem{b} \subseteq \sem{a}$. Therefore, $P(b) = P(a)$. \\[1mm]
For the converse, assume $P$ is both \EXT{} and \SIND{}. Define the following relation $x \succ_P y$
\[ x \succ_P y \equiv \forall a \in T X (x, y \in \sem{a} \to P(a) \neq y) \]
In particular, this implies that $x$ is chosen in $a \in TX$ when $\sem{a} = \{x, y\}$. Let us first show that this relation is actually a preference relation. In order to see that it is \emph{transitive}, assume $x \succ_P y$ and $y \succ_P z$, and fix $a \in TX$ such that $x, z \in \sem{a}$. We need to show that $P(a) \neq z$. Let $a'$ be any problem which has $a$ as a sub-problem, but any other sub-problem or immediate value only has $y$. By $y \succ_P z$, we have that $P(a') \neq z$. But, by $x \succ_P y$, since $x, z \in \sem{a'}$, we also have that $P(a') \neq y$.  But since $P$ is inductive, and by the way $a'$ was constructed, we must have that $P(a') = P(a)$. Therefore, $P(a) \neq z$. In order to show that this relation is also \emph{total}, suppose there were $x, y$ such that $\neg (x \succ_P y)$ and $\neg (y \succ_P x)$. That implies that for some $a$ and $b$, both containing $x$ and $y$, we have that $P(a) = x$ and $P(b) = y$. Let $C = \sem{a} \cap \sem{b}$, which is guaranteed to be non-empty since both $a$ and $b$ contain $x$ and $y$. Let $a'$ be such that $\sem{a'} = \sem{a}$ but $a'$ has a sub-problem $c$ such that $\sem{c} = C$, and the only place where $x,y$ appear in $a'$ is in the sub-problem $c$. Similarly, we can also find a $b'$ such that $\sem{b'} = \sem{b}$ but $b'$ also has $c$ as a sub-problem, with the same property that $x, y$ only appear in $c$. Since $P$ is extensional, we have that $P(a) = P(a')$ and $P(b) = P(b')$. But since $P$ is also inductive, we must have that $P(a') = P(c)$ and  $P(b') = P(c)$, which is a contradiction since $P(a) = x \neq y = P(b)$. That concludes the proof that $x \succ_P y$ is a preference relation. It remains to see that $P(a) = \max_{\succ_P} \sem{a}$. But that is indeed the case since $x \succ_P y$ implies that $y$ is never chosen when $x$ is present. If $y = P(a)$ was not the maximal element of $\sem{a}$ according to this order, there would be a $x \in \sem{a}$ with $x \succ_P y$, meaning that $y$ could not have been chosen in $a$, contradiction. \hfill $\Box$ \\

Proof of Theorem \ref{theorem:vmain}:

{\bf Proof}. A simple relativization of the proof of Theorem \ref{theorem:main}. Assume $P$ is ${\cal V}$-rationalizable and let $c \colon 2^X \to X$ be a rationalizable choice function such that
\begin{itemize}
	\item[(i)] $P(a) = c(\sem{a})$, for all $a \in {\cal V}$.
\end{itemize}
Let $(X, \succ)$ be the \emph{strict preference relation} underpinning $c$, i.e.
\begin{itemize}
	\item[(ii)] $c(A) = \{x \in A \; | \; \forall y \in A (x \neq y \to x \succ y)\}$.
\end{itemize}
In order to show \EXT{}, let $a, b \in {\cal V}$ be two represented problems such that $\sem{a} = \sem{b}$. Clearly, $c(\sem{a}) = c(\sem{b})$, and hence, by (i), $P(a) = P(b)$.  In order to show that $P$ is also \SIND{}, let $a \in {\cal V}$ be a fixed represented problem, and assume $P(a) \not\in \ival(a)$. Since $P(a) \in \sem{a}$ there must be a $b \in \spro(a)$ such that $P(a) \in \sem{b}$. By (ii), $P(a)$ is a maximal element in $\sem{a}$, and hence also in $\sem{b}$, since $\sem{b} \subseteq \sem{a}$. Therefore, $P(b) = P(a)$. \hfill $\Box$ \\

Proof of Theorem  \ref{theorem:vmain2}:

{\bf Proof}. Again, a simple relativization of the proof of Theorem \ref{theorem:main}. Let ${\cal V}$ be expressive and $T$-closed, and assume $P$ is both ${\cal V}$-\EXT{} and ${\cal V}$-\SIND{}. Define the following relation $x \succ_P y$
\[ x \succ_P y \equiv \forall a \in {\cal V} (x, y \in \sem{a} \to P(a) \neq y) \]
Let us first show that this relation is actually a preference relation. In order to see that it is \emph{transitive}, assume $x \succ_P y$ and $y \succ_P z$, and fix $a \in {\cal V}$ such that $x, z \in \sem{a}$. We need to show that $P(a) \neq z$. Let $a' \in {\cal V}$ be any problem which has $a$ as a sub-problem, but any other sub-problem or immediate value only has $y$. Such $a'$ exists under the assumption that ${\cal V}$ is $T$-closed. By $y \succ_P z$, we have that $P(a') \neq z$. But, by $x \succ_P y$, since $x, z \in \sem{a'}$, we also have that $P(a') \neq y$.  But since $P$ is inductive, and by the way $a'$ was constructed, we must have that $P(a') = P(a)$. Therefore, $P(a) \neq z$. In order to show that this relation is also \emph{total}, suppose there were $x, y$ such that $\neg (x \succ_P y)$ and $\neg (y \succ_P x)$. That implies that for some $a, b \in {\cal V}$, both containing $x$ and $y$, we have that $P(a) = x$ and $P(b) = y$. Let $C = \sem{a} \cap \sem{b}$, which is guaranteed to be non-empty since both $a$ and $b$ contain $x$ and $y$. Let $a'$ be such that $\sem{a'} = \sem{a}$ but $a'$ has a sub-problem $c$ such that $\sem{c} = C$, and the only place where $x,y$ appear in $a'$ is in the sub-problem $c$. Such $c'$ exists under the assumption that ${\cal V}$ is expressive ($c \in {\cal V}$ exists with $\sem{c} = C$) and $T$-closed ($c' \in {\cal V}$ exists with $c$ as a sub-problem). Similarly, we can also find a $b'$ such that $\sem{b'} = \sem{b}$ but $b'$ also has $c$ as a sub-problem, with the same property that $x, y$ only appear in $c$. Since $P$ is extensional, we have that $P(a) = P(a')$ and $P(b) = P(b')$. But since $P$ is also inductive, we must have that $P(a') = P(c)$ and  $P(b') = P(c)$, which is a contradiction since $P(a) = x \neq y = P(b)$. That concludes the proof that $x \succ_P y$ is a preference relation. It remains to see that $P(a) = \max_{\succ_P} \sem{a}$. But that is indeed the case since $x \succ_P y$ implies that $y$ is never chosen when $x$ is present. If $y = P(a)$ was not the maximal element of $\sem{a}$ according to this order, there would be a $x \in \sem{a}$ with $x \succ_P y$, meaning that $y$ could not have been chosen in $a$, contradiction. \hfill $\Box$ \\

Proof of Theorem \ref{thm-ue}:

{\bf Proof}. From left-to-right this is trivial since when ${\cal V}$ uniquely
determines the representation of a set $A \subseteq X$, then any two $a_1, a_2
\in T X$ such that $\sem{a_1} = \sem{a_2}$ will be such that $a_1 = a_2$, and
hence $P(a_1) = P(a_2)$. For the other direction, suppose all procedures $P$ are
${\cal V}$-\EXT{}. Assume ${\cal V}$ is not uniquely expressive. There must exist $a_1, a_2 \in TX$ such that $A = \sem{a_1} = \sem{a_2}$ but $a_1 \neq a_2$. Let $x, y$ be two distinct elements of $A$, and define a procedure $P$ such that $P(a_1) = x$ but $P(a_2) = y$. This is always possible because procedures are able to inspect not just the values of $a_i$ but the structure of the representation. But this is a contradiction because $P$ is by assumption ${\cal V}$-\EXT{}. \hfill $\Box$ \\

Proof of Corollary \ref{cor-unique}:

{\bf Proof}. Assume every $P$ on $T X$ is either ${\cal V}$-rational or ${\cal V}$-\EXT{}, but suppose ${\cal V}$ is not uniquely expressive. By Theorem \ref{thm-ue} there exists a procedure $P \colon TX \to X$ which is not ${\cal V}$-\EXT{}. However, by Theorem \ref{theorem:vmain}, $P$ is also not ${\cal V}$-rationalizable, which is a contradiction. \hfill $\Box$ \\

Proof of Theorem \ref{main-theorem2}:

{\bf Proof}. Since $P \colon TX \to X$ is intensional, we know that for any
given equivalence class ${\bf a} \in \Tquotient$ the procedure $P$ always
chooses the values on the same index $i$, i.e. $P(a) = a[i]$ for all $a \in {\bf
  a}$. Let $A \subseteq X$ be fixed and $a \in {\cal V}$ be such that $\sem{a} =
A$. Such $a \in {\cal V}$ exists since we assume ${\cal V}$ is expressive. Let
$v = P(a)$. We claim that for any other $b \in {\cal V} \cap {\bf a}$ with
$\sem{b} = A$ we also have $b[i] = v$. Indeed, if $b \in {\cal V} \cap {\bf a}$
we have that $P(b) = b[i]$, but also $P(a) = a[i]$. By the assumption that $P$ is
${\cal V}$-rational, we must have that $b[i] = a[i] = \max_{\succ} A$, for some strict preference relation $(X, \succ)$, which implies $P(b) = b[i] = a[i] = v$. \hfill $\Box$ \\

\end{document}